\documentclass[%
prl,
 amsmath,amssymb,
 reprint,%
pr]{revtex4-1}

\usepackage{bm}
\usepackage[colorlinks]{hyperref}
\usepackage {graphicx}
\usepackage {amsmath}
\usepackage {hyperref}
\usepackage{upgreek}
\usepackage{gensymb}

\def\gev{GeV/$c^2$}
\def\mev{MeV/$c^2$}

\begin{document}


\title{Directional Sensitivity In Light-Mass Dark Matter Searches With Single-Electron Resolution Ionization Detectors}


\author{Fedja Kadribasic, Nader Mirabolfathi}
\affiliation{Department of Physics and Astronomy, Texas A\&M University}

\author{Kai Nordlund, Andrea E. Sand, E.\ Holmstr{\"o}m, Flyura Djurabekova}
\affiliation{Helsinki Institute of Physics and Department of Physics, University of Helsinki}


\date{\today}

\begin{abstract}

We propose a method using solid state detectors with directional sensitivity to dark matter interactions to detect low-mass Weakly Interacting Massive Particles (WIMPs) originating from galactic sources.\  In spite of a large body of literature for high-mass WIMP detectors with directional sensitivity, no available technique exists to cover WIMPs in the mass range $<$1~\gev.\ We argue that single-electron resolution semiconductor detectors allow for directional sensitivity once properly calibrated.\ We examine commonly used semiconductor material response to these low-mass WIMP interactions.\end{abstract}


\maketitle



Many astrophysical observations indicate that standard model particles compose only 15\% of the matter in the universe~\cite{Ade:2013zuv}.\ Understanding the nature of 
dark matter, the remaining 85\%, is of fundamental importance to cosmology, astrophysics, and high energy particle physics.\ Although Weakly Interacting Massive Particles (WIMPs) of mass 10-100 \gev ~have been the main interest of the majority of direct dark matter detection experiments, recent signal claims, compelling theoretical models, and the lack of a convincing signal at those masses have shifted the old paradigm to include broader regions in the dark matter parameter space well below 10 \gev \cite{CF1}.

Direct detection experiments attempt to detect WIMPs via their elastic interaction with detector nuclei~\cite {Gaitskell:2004gd}.\ Since very low energy nuclear recoils and small interaction rates from these low-mass WIMPs are expected, large-mass detectors with very low threshold are desirable.\ 
Solid state detectors, especially those utilizing phonon-mediated readout technology, have already reached the sensitivities required to detect these very-low-mass WIMPs or are braced to do so~\cite{Agnese:2013lua}.

Both reducible (environmental) and irreducible (solar neutrino) backgrounds that may mimic WIMPs affect WIMP direct search experiment sensitivity.\ A potential tool to circumvent these backgrounds is the directionality of the WIMPs' signal due to Earth's motion through their isothermal halo distribution in our galaxy.\ The WIMP velocity distribution in the lab frame, and hence the expected direction of the WIMP-induced recoils, varies daily depending on the angular orientation of the detectors with respect to the galactic WIMP flux.

Although many experiments propose to track WIMP-induced recoils using low-pressure gas or even liquid scintillators, they do not offer low enough energy thresholds to detect recoils from low-mass WIMP interactions ($<$1~\gev ) \cite{DirectionalReview}.\ Furthermore, low-pressure-gas detectors require prohibitively large volumes to detect any WIMP signal.\ 
We argue that single-electron resolution phonon-mediated semiconductor detectors, such as those in development for SuperCDMS and future generation-3 dark matter experiments, are sensitive to the nuclear recoil direction and can be used for a directional dark matter search.\ Our method uses the fundamental processes involved in nuclear recoil ionization excitation whose threshold exhibits a strong recoil direction dependence.\ Recent progress on phonon-mediated detectors, especially Neganov-Luke phonon amplification detectors~\cite{Luke_CDMSlite}, promises future large-mass semiconductor detectors with single-electron resolution~\cite{Contactfree}.





Neither experimental data nor an established computational framework exists
to estimate the minimum energy required to create single electron-hole pair excitations via nuclear recoil interactions.
Based on two recent observations, we assume that this so-called ionization threshold correlates with crystallographic orientation in the direction of the nuclear recoil.
Firstly, strong experimental and theoretical evidence indicates that the ionization threshold, often referred to as electronic stopping, displays a nonlinear dependence on projectile velocity at low projectile energies due to electronic band structure effects~\cite{Val03,Mar09,Pri12}.
Secondly, recent time-dependent density functional theory (TDDFT) calculations demonstrate the appearance
of an intermediate band gap state for self-recoils in silicon (Si) that arises when the projectile occupies an interstitial position, which 
serves to modulate the sharp 
ionization threshold 
in insulators~\cite{Lim16}. This intermediate ``electron elevator''~\cite{Lim16} state enables excitations across the band gap even when the energy transfer in ion-electron collisions remains below the level needed for a direct transition from the valence to conduction band~\cite{Hor16}.
Electronic excitation is thus observed for projectiles with velocities as low as 0.1 \AA/fs~\cite{Lim16} corresponding to an ionization threshold below 15 eV for an Si projectile.
Because this defect state exists due to the interstitial atom configuration, and the energy level oscillates as a function of the position of the interstitial, 
the effective ionization threshold 
should depend on the recoil angle. This energy is comparable to the directionally sensitive threshold displacement energy (TDE), \textit{i.e.} the minimum energy required to eject the recoiling nucleus permanently to a crystal defect position. 
Hence the recoil trajectory, and 
the probability of an atom reaching an interstitial position
to 
facilitate  
electron-hole pair excitation,
should depend on the recoil angle. 
We model the variations in the energy landscape experienced by low-energy recoils via the TDE.

We consider the threshold variation for two common detector materials, Ge and Si.\  For both, density-functional theory (DFT) molecular dynamics (MD) simulations have previously obtained the average threshold displacement energy and the direction-specific values in the $\langle100\rangle$ and $\langle111\rangle$ crystal directions \cite{Hol08a,Hol10a}.\ 

To determine the full TDE surface to high statistical accuracy,
we follow the procedure described in Ref. \cite{Nor05c} with
tens of thousands of different recoil directions.
Put succinctly, a 4096 atom Ge or Si simulation cell was equilibrated at 0.04 K (an upper limit for the experimental detector temperature), giving all atoms random thermal displacements.\ After this, an atom was randomly chosen within the central eight unit cells of the simulation cell and given a recoil of energy $E$ in a randomly selected direction $(\theta,\phi)$ in three dimensions, where $\theta$ is defined as the polar angle off the [001] crystal direction and $\phi$ as the azimuthal angle from the [100] direction towards [010].\ The evolution of the collision sequence thus generated was simulated for 10 ps,
and we analyze possible defect creation automatically using Wigner-Seitz and potential energy criteria \cite{Nor05c}.\ For each atom and direction, the energy $E$ was increased from 2 eV in steps of 1~eV until a stable defect was created.

The outcome of MD simulations depends crucially on the interatomic potential used \cite{Allen-Tildesley,Nor05c}.\  Hence, for the purpose of this study, we compared several different Ge and Si interatomic potentials with the DFT results.\  Among the three tested interatomic potentials for Ge \cite{Din86,Nor97f,Pos09}, the modified Stillinger-Weber (SW) potential from Ref.\ \cite{Nor97f} reproduced all of the reported DFT threshold displacement energies \cite{Hol10a} within the error bars, giving us high confidence of a reliable description of the entire data range.\ Hence, this potential was used for all Ge simulations.\ We have previously shown that, out of three commonly used Si potentials, SW \cite{Sti85} reproduces the DFT and experimental results the best.\ Consequently, we use this potential to calculate the rates in Si.

\begin{figure}
  \begin{center} \includegraphics[width=0.98\columnwidth]{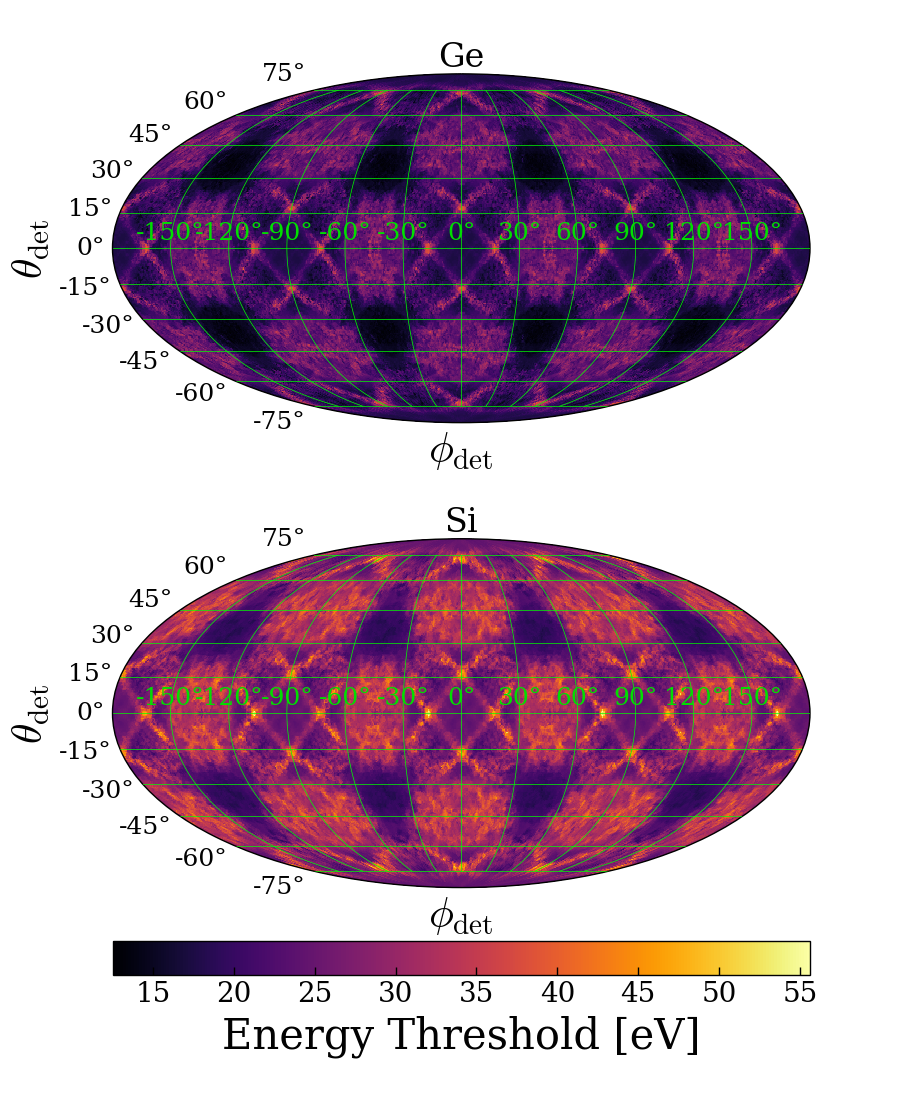} \end{center}
\caption{\label{thresh}
Threshold displacement energy surface in different crystal directions in Ge (top) and Si (bottom) determined from classical MD simulations illustrated with a Mollweide projection.\ These plots represent the averages over the eight threshold surface datasets.\ 
Darker regions correspond to a lower energy threshold and, hence, a higher differential rate (see Fig.~\ref {ang}).\ 
}

\end{figure}

In total, we simulate about 85,000 directions for Ge and about 24,000 for Si a total of eight times.\ Fig.~\ref{thresh} illustrates the average over the resulting threshold displacement energy surfaces for Ge and Si.\ The symmetry of the diamond crystal structure causes the periodicity with respect to $\phi = 45^\circ$, and the zero-point quantum motion of atoms in the lattice causes the graininess in the plots.\ Fig.\ \ref{thresh} shows that the energy threshold to create a defect strongly depends on the nuclear recoil direction.\ The Ge threshold ranges from 12.5 eV to 63.5 eV whereas that for Si ranges from 17.5 eV to 63.5 eV.\ 

The expected total WIMP signal rate above the detection threshold can be calculated by integrating the differential rate over the recoil angle and recoil energy.\ In the case of a charge detector, assuming that defect and electronic excitation thresholds are equal, the energy thresholds, henceforth referred to as $E_{th} (\theta, \phi)$ and shown in Fig.\ \ref{thresh}, simply provide the lower limit to the integral
\begin {equation} \label {integral} R(t) = \oint_{4 \pi} \int_{E_{th} (\theta, \phi)}^{E_r^{max}} \dfrac {\partial^2 R} {\partial E_r \partial \Omega_r} \text{d} E_r \text{d} \Omega_r.\ \end {equation}
This rate, measured by a fixed detector on the surface of Earth, which is moving and rotating relative to the WIMP halo, should, therefore, exhibit a diurnal modulation since $E_{th}$ is a function of $\theta$ and $\phi$.\ Below, we describe our procedure to calculate this integral.\ 

~\cite {rate} gives the integrand in Eq.~\ref {integral}, the differential interaction rate between halo WIMPs and detectors for spin-independent interactions, as
\begin {multline} \label {dmrate} \dfrac {\partial^2 R} {\partial E_r \partial \Omega_r} = \dfrac {\rho_0 \sigma_{\chi-n} A^2} {4 \pi m_\chi \mu_{\chi n}^2} \times F^2 (E_r) \hat {f}_{\text {lab}} (v_{\text {min}}, \bm{\hat {q}_r}; t) \end {multline}
where $m_\chi$ is the WIMP mass, $\mu_{\chi n}$ is the WIMP-nucleon reduced mass, $\rho_0 = 0.3\ \text {GeV cm}^{-3}$ is the local dark matter density, $A$ is the mass number of the nucleus, $\sigma_{\chi-n}$ is the WIMP-nucleon cross section, $v_{\text {min}} = \sqrt {2 m_N E_r} / 2 \mu_{\chi n}$ is the minimum WIMP speed required to produce a nuclear recoil of energy $E_r$ for a given nuclear mass $m_N$, and $F^2 (E_r)$ is the Helm nuclear form factor~\cite {fsq}.

~\cite {rate} gives the Radon transform of the WIMP velocity distribution as
\begin {multline} \hat {f}_{\text {lab}}  (v_{\text {min}}, \bm{\hat {q}}; t) = \dfrac {1} {N_{\text {esc}} \sqrt {2 \pi \sigma_\nu^2}} \times \\
\left[ \text {exp} \left(-\dfrac {|v_{\text {min}} + \bm {\hat {q}} \cdot \bm{v}_{\text {lab}}|^2} {2 \sigma_\nu^2}\right)  - \text {exp} \left(-\dfrac {v_{\text{esc}}^2} {2 \sigma_\nu^2} \right) \right] \end {multline}
where $\bm {\hat {q}}$ is the recoil direction in detector coordinates, $\bm {v}_{\text {lab}}$ is the velocity of the laboratory relative to a stationary observer, $v_{\text {esc}}$ is the circular escape velocity at the Solar System's distance from the Milky Way's center, $\sigma_v = v_0 / \sqrt(2)$ is the dark matter velocity dispersion, and $N_{\text {esc}}$ is a normalization factor.\ We use $v_0 = 220\ \text {km s}^{-1}$ for the circular speed and $v_{\text {esc}} = 544\ \text {km s}^{-1}$~\cite {rate}.

\begin {figure}
\centering \includegraphics [width = 0.98 \columnwidth] {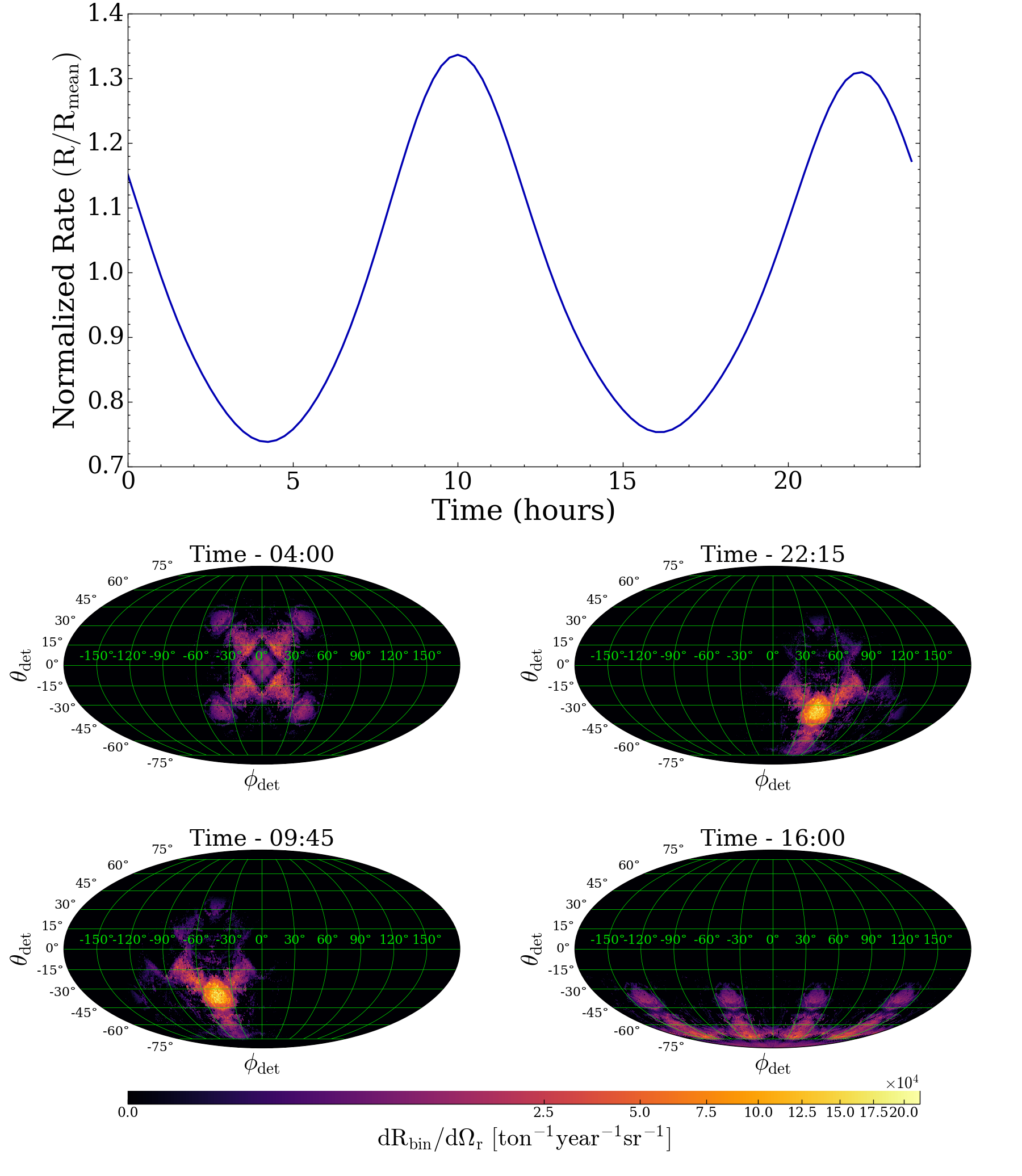}
\caption {(Top) Normalized integrated rate with respect to mean over one day for a 300 \mev\ WIMP at the SNOLAB site.\ (Bottom) Angular distribution of  differential rate per steradian for a nucleon cross section of $10^{-39} \text {cm}^2$ over one day for a 300 \mev\ WIMP at the SNOLAB site.\ Each angle plot corresponds to a local extremum of the integrated rate.\ }
\label {ang}
\end {figure}

Following Appendix B of Ref.~\cite {vel}, we find the total lab velocity using the contributions due to galactic rotation, solar motion, Earth's revolution, and Earth's rotation.\ The calculations assume a detector at SNOLAB coordinates $(46.4719 \degree, 81.1868 \degree)$.\  The variation in lab-frame speed of the dark matter gives a $\sim$6\% annual and nearly negligible diurnal modulation \footnote{Earth's revolution around the sun causes the annual modulation, whereas Earth's rotation causes the daily one.}.

We calculate signal rates assuming a detector with~1~eV resolution, 100\% detection efficiency, and no backgrounds.\  We perform the integral in Eq.~\ref{integral} over the recoil energy $E_r$ and recoil angle $\Omega_r$ using 48 time steps on September 6, 2015.\ The date was chosen to cross-check our differential rate calculations with those in Ref.~\cite{rate}.\ An equidistant coordinate partition interpolation of the data shown in Fig.~\ref{thresh} is performed on a grid with 2400 elements in the $\theta$ direction and 4800 in the $\phi$ direction.\ For faster computation, the grid is resampled to a size of 196,608 pixels using the HEALPix algorithm~\cite {hp}.\  We compute a multidimensional Riemann sum over each dimension with 200 sample points for $E_r$ and 196,608 for $\Omega_r$.

Fig.~\ref {ang} shows the integrated event rate for a WIMP of mass 300~\mev~and cross section~$\sigma_\text{WIMP-nucleon}$=$\text10^{-39}  \text{cm}^{2}$~over the course of one day (Sept 6, 2015).\ The mass and cross section were arbitrarily chosen within the unexplored region in the halo WIMP parameter space.\ Also shown in this figure are 
 the angular distributions of the rates at four different times illustrating recoil orientation change with respect to the crystal over the course of the day.\ 
As Earth rotates, more events are detected at the energy minima than the maxima, which leads to an integrated rate modulation (in this case~$\sim$60$\%$) with a phase imposed by the threshold data in Fig.~\ref {thresh}.

\begin {figure}
\centering \includegraphics [width = 0.98 \columnwidth] {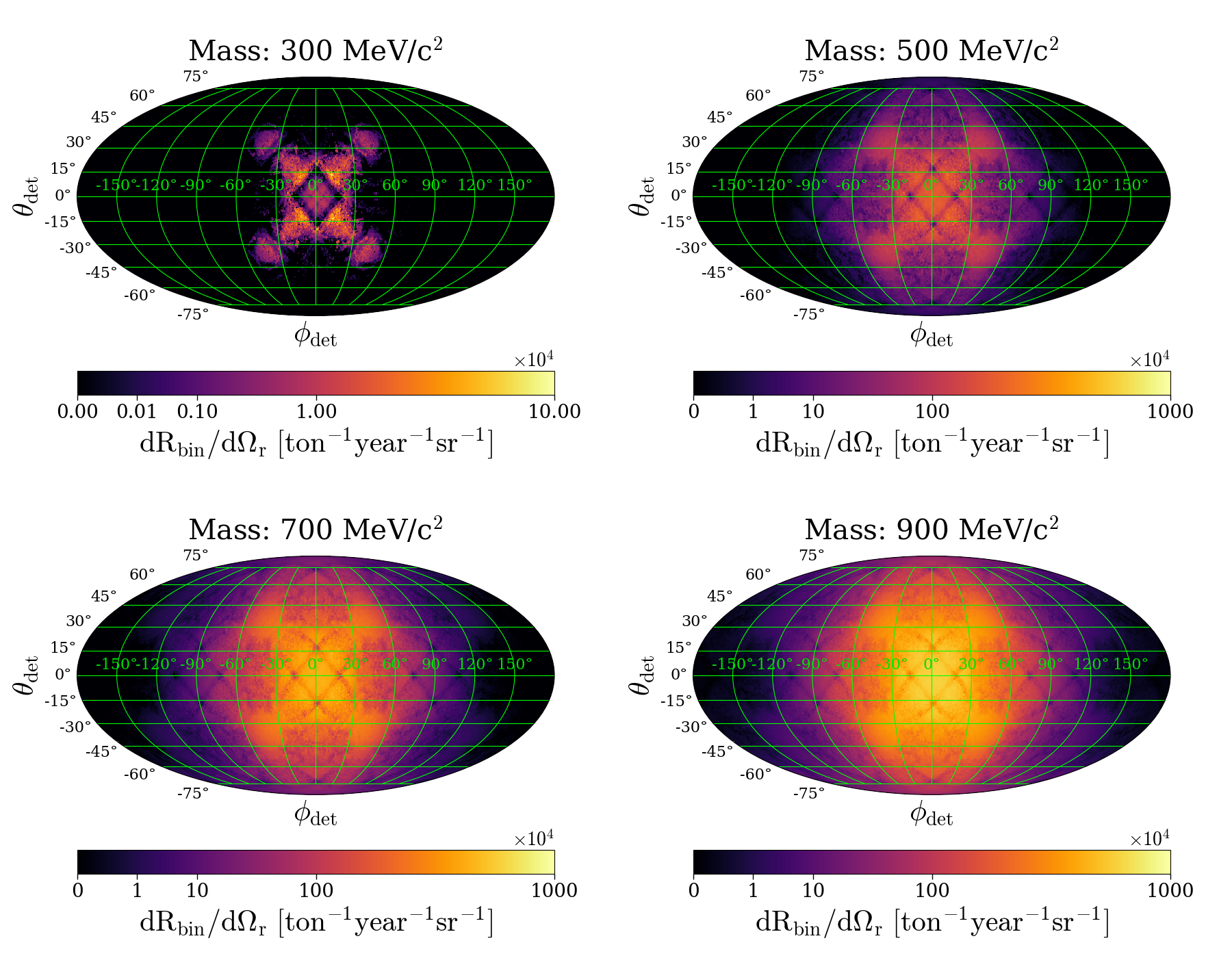}
\caption {Angular distribution of differential rate per steradian for a Ge detector assuming a nucleon cross section of $10^{-39} \text {cm}^2$ for several WIMP masses at 4:00 on September 6, 2015.\ As the WIMP mass increases, the differential rate angular spread increases due to the Maxwell-Boltzmann velocity distribution and hard-sphere scattering acting in conjunction with the energy thresholds (see Fig.~\ref {thresh}).}
\label {angmass}
\end {figure}

We repeated this study for WIMPs covering a mass range between 230~\mev~and 10~\gev~in Ge and between 165 \mev~and 10 \gev~in Si.\ Lighter-mass WIMPs do not produce stable defects or electron-hole pair excitation even when traveling at the escape velocity $v_{\text {esc}} = 544\ \text {km s}^{-1}$.\ Fig.~\ref{angmass} shows the recoil angular distribution in Ge at a given time (4:00 on September 6, 2015) for a sample of WIMP masses in this range.\ As shown in this figure, larger mass WIMPs produce a broader recoil angle distribution.\ Hence, the integrated signal rate associated with larger mass WIMPs is less sensitive to the crystallographic orientation of the detector.\ We expect smaller event rate modulation for larger mass WIMPs due to this effect.

\begin {figure}
\centering \includegraphics [width = 0.98 \columnwidth] {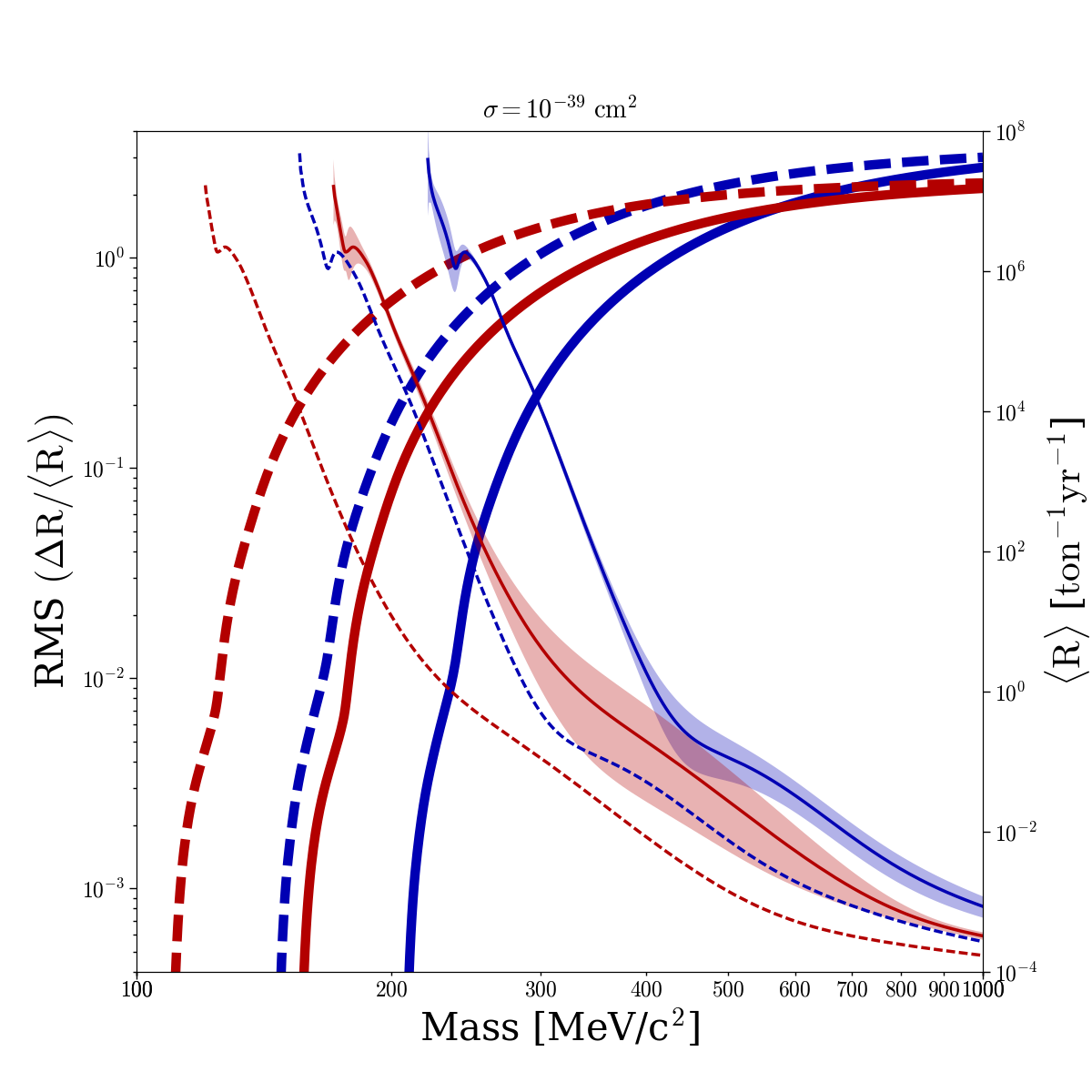}
\caption {Normalized RMS of the rate modulation (
left axis, thin lines) 
and mean rate (right axis, thick lines) as a function of dark matter mass for Ge (blue) and Si (red).\ 
A WIMP-nucleon cross section of $10^{-39} \text {cm}^2$ is assumed.\ Normalized RMS modulation error is given by the shaded regions.\ Mean rate error is negligible and consequently not included.\ The thick and thin dashed curves show the normalized RMS modulation and mean rate given thresholds half of those used for the solid curves.\ 
}
\label {rms}
\end {figure}

To assess the strength of the signal rate modulation with respect to the signal mean rate, we perform a normalized root-mean squared (RMS) modulation integral over one day
\begin {equation} 
\label {rmseq} R_{\text {RMS, norm}} = \sqrt {\dfrac {1} {\langle R \rangle^2 \Delta t} \oint_{\Delta t} (R(t) - \langle R \rangle)^2 dt}
\end {equation}
where $\langle R \rangle$ is the average value over $\Delta t$, which is one solar day (24 hours).\ The results of these studies are shown in Fig.~\ref {rms}.\ We find a clear rate modulation for WIMPs of mass below 1 \gev.\ As expected, while the signal mean rate (thicker graph) decreases at lower WIMP masses, the modulation gains strength
, which enables the experiments to maintain their signal to background ratio by only looking at the time intervals when the signal rate is maximized.\ 
Furthermore, since the Si nucleus is less massive than that of Ge, the energy transfer from a WIMP is more efficient; hence, a lower WIMP mass is required to transfer recoil energy sufficient to overcome the threshold displacement energy.\ Consequently, the peak of the modulation appears at lower WIMP masses for Si than for Ge.



The stochastic threshold displacement caused by the zero-point quantum motion of atoms was included based on the Debye model, which allows calculating the one-dimensional RMS displacement amplitude~\cite {Gemmell:1974ub, Debyedisplacements}.
We calculate eight separate threshold datasets for Ge and Si using MD simulations.\ In Fig.~\ref {rms}, the RMS curves and shaded regions show the mean and standard deviation of the normalized RMS modulation values over all eight datasets.\ 
The kinks in the normalized RMS modulation curves correspond to the various length-scale transitions in the energy threshold shown in Fig.~\ref {thresh}, which reveal themselves due to the larger solid angle coverage at higher dark matter masses.\ 

We reproduce the normalized RMS modulation and mean rate using energy thresholds 50\% of those in Fig.~\ref {thresh} as dashed curves.\ As expected, there is a clear diurnal modulation, albeit at lower masses.\ This work provides strong motivation for experimental validation of the energy thresholds for ionization excitations via nuclear elastic scattering in Ge or Si.\ 

Based on the substantiated evidences for the threshold dependence on the nuclear recoil direction, we project a strong diurnal modulation in the expected detection rate of galactic halo WIMPs.\ This modulation strongly depends on the target detector material and WIMP mass, and, together with the overall mean rate, it provides an extra handle to determine WIMP mass and cross section independently.\ This effect can be used to discriminate WIMPs from solar neutrino backgrounds that will become the irreducible background for all dark matter search experiments.\ Even if future experiments find different ionization thresholds, the anisotropy predicted for electron-hole pair creation could still cause modulation in dark matter signal, albeit over a different mass range.\ The significance of these results motivates thorough semiconductor detector calibration at low recoil energies.

N.\ M.\ acknowledges Mitchell Institute For Fundamental Physics financial support.\ E.\ H.\ acknowledges financial support from the Emil Aaltonen foundation
and the Academy of Finland through the Centres of Excellence Program
(Project No.\ 251748).




\bibliographystyle{apsrev4-1}
\bibliography{Diurnal_DM_v02_FK_No_Red_2018_03_19}

\begin{thebibliography}{27}%
\makeatletter
\providecommand \@ifxundefined [1]{%
 \@ifx{#1\undefined}
}%
\providecommand \@ifnum [1]{%
 \ifnum #1\expandafter \@firstoftwo
 \else \expandafter \@secondoftwo
 \fi
}%
\providecommand \@ifx [1]{%
 \ifx #1\expandafter \@firstoftwo
 \else \expandafter \@secondoftwo
 \fi
}%
\providecommand \natexlab [1]{#1}%
\providecommand \enquote  [1]{``#1''}%
\providecommand \bibnamefont  [1]{#1}%
\providecommand \bibfnamefont [1]{#1}%
\providecommand \citenamefont [1]{#1}%
\providecommand \href@noop [0]{\@secondoftwo}%
\providecommand \href [0]{\begingroup \@sanitize@url \@href}%
\providecommand \@href[1]{\@@startlink{#1}\@@href}%
\providecommand \@@href[1]{\endgroup#1\@@endlink}%
\providecommand \@sanitize@url [0]{\catcode `\\12\catcode `\$12\catcode
  `\&12\catcode `\#12\catcode `\^12\catcode `\_12\catcode `\%12\relax}%
\providecommand \@@startlink[1]{}%
\providecommand \@@endlink[0]{}%
\providecommand \url  [0]{\begingroup\@sanitize@url \@url }%
\providecommand \@url [1]{\endgroup\@href {#1}{\urlprefix }}%
\providecommand \urlprefix  [0]{URL }%
\providecommand \Eprint [0]{\href }%
\providecommand \doibase [0]{http://dx.doi.org/}%
\providecommand \selectlanguage [0]{\@gobble}%
\providecommand \bibinfo  [0]{\@secondoftwo}%
\providecommand \bibfield  [0]{\@secondoftwo}%
\providecommand \translation [1]{[#1]}%
\providecommand \BibitemOpen [0]{}%
\providecommand \bibitemStop [0]{}%
\providecommand \bibitemNoStop [0]{.\EOS\space}%
\providecommand \EOS [0]{\spacefactor3000\relax}%
\providecommand \BibitemShut  [1]{\csname bibitem#1\endcsname}%
\let\auto@bib@innerbib\@empty
\bibitem [{\citenamefont {Ade}\ \emph {et~al.}(2014)\citenamefont {Ade} \emph
  {et~al.}}]{Ade:2013zuv}%
  \BibitemOpen
  \bibfield  {author} {\bibinfo {author} {\bibfnamefont {P.~A.~R.}\
  \bibnamefont {Ade}} \emph {et~al.} (\bibinfo {collaboration} {Planck}),\
  }\href {\doibase 10.1051/0004-6361/201321591} {\bibfield  {journal} {\bibinfo
   {journal} {Astron. Astrophys.}\ }\textbf {\bibinfo {volume} {571}},\
  \bibinfo {pages} {A16} (\bibinfo {year} {2014})},\ \Eprint
  {http://arxiv.org/abs/1303.5076} {arXiv:1303.5076 [astro-ph.CO]} \BibitemShut
  {NoStop}%
\bibitem [{\citenamefont {Cushman}\ \emph {et~al.}(2013)\citenamefont {Cushman}
  \emph {et~al.}}]{CF1}%
  \BibitemOpen
  \bibfield  {author} {\bibinfo {author} {\bibfnamefont {P.}~\bibnamefont
  {Cushman}} \emph {et~al.},\ }\bibfield  {booktitle} {\emph {\bibinfo
  {booktitle} {{Proceedings, 2013 Community Summer Study on the Future of U.S.
  Particle Physics: Snowmass on the Mississippi (CSS2013): Minneapolis, MN,
  USA, July 29-August 6, 2013}}},\ }\href
  {https://inspirehep.net/record/1262767/files/arXiv:1310.8327.pdf} {\
  (\bibinfo {year} {2013})},\ \Eprint {http://arxiv.org/abs/1310.8327}
  {arXiv:1310.8327 [hep-ex]} \BibitemShut {NoStop}%
\bibitem [{\citenamefont {Gaitskell}(2004)}]{Gaitskell:2004gd}%
  \BibitemOpen
  \bibfield  {author} {\bibinfo {author} {\bibfnamefont {R.~J.}\ \bibnamefont
  {Gaitskell}},\ }\href {\doibase 10.1146/annurev.nucl.54.070103.181244}
  {\bibfield  {journal} {\bibinfo  {journal} {Ann. Rev. Nucl. Part. Sci.}\
  }\textbf {\bibinfo {volume} {54}},\ \bibinfo {pages} {315} (\bibinfo {year}
  {2004})}\BibitemShut {NoStop}%
\bibitem [{\citenamefont {Agnese}\ \emph {et~al.}(2014)\citenamefont {Agnese}
  \emph {et~al.}}]{Agnese:2013lua}%
  \BibitemOpen
  \bibfield  {author} {\bibinfo {author} {\bibfnamefont {R.}~\bibnamefont
  {Agnese}} \emph {et~al.} (\bibinfo {collaboration} {SuperCDMS}),\ }\href
  {\doibase 10.1103/PhysRevLett.112.041302} {\bibfield  {journal} {\bibinfo
  {journal} {Phys. Rev. Lett.}\ }\textbf {\bibinfo {volume} {112}},\ \bibinfo
  {pages} {041302} (\bibinfo {year} {2014})},\ \Eprint
  {http://arxiv.org/abs/1309.3259} {arXiv:1309.3259 [physics.ins-det]}
  \BibitemShut {NoStop}%
\bibitem [{\citenamefont {Mayet}\ \emph {et~al.}(2016)\citenamefont {Mayet}
  \emph {et~al.}}]{DirectionalReview}%
  \BibitemOpen
  \bibfield  {author} {\bibinfo {author} {\bibfnamefont {F.}~\bibnamefont
  {Mayet}} \emph {et~al.},\ }\href {\doibase 10.1016/j.physrep.2016.02.007}
  {\bibfield  {journal} {\bibinfo  {journal} {Phys. Rept.}\ }\textbf {\bibinfo
  {volume} {627}},\ \bibinfo {pages} {1} (\bibinfo {year} {2016})},\ \Eprint
  {http://arxiv.org/abs/1602.03781} {arXiv:1602.03781 [astro-ph.CO]}
  \BibitemShut {NoStop}%
\bibitem [{\citenamefont {Luke}\ \emph {et~al.}(1990)\citenamefont {Luke},
  \citenamefont {Beeman}, \citenamefont {Goulding}, \citenamefont {Labov},\
  and\ \citenamefont {Silver}}]{Luke_CDMSlite}%
  \BibitemOpen
  \bibfield  {author} {\bibinfo {author} {\bibfnamefont {P.~N.}\ \bibnamefont
  {Luke}}, \bibinfo {author} {\bibfnamefont {J.}~\bibnamefont {Beeman}},
  \bibinfo {author} {\bibfnamefont {F.~S.}\ \bibnamefont {Goulding}}, \bibinfo
  {author} {\bibfnamefont {S.~E.}\ \bibnamefont {Labov}}, \ and\ \bibinfo
  {author} {\bibfnamefont {E.~H.}\ \bibnamefont {Silver}},\ }\bibfield
  {booktitle} {\emph {\bibinfo {booktitle} {{Nucl. Instrum. Methods A289 (1990)
  406-409.}}},\ }\href {\doibase 10.1016/0168-9002(90)91510-I} {\bibfield
  {journal} {\bibinfo  {journal} {Nucl. Instrum. Meth.}\ }\textbf {\bibinfo
  {volume} {A289}},\ \bibinfo {pages} {406} (\bibinfo {year}
  {1990})}\BibitemShut {NoStop}%
\bibitem [{\citenamefont {Mirabolfathi}\ \emph {et~al.}(2017)\citenamefont
  {Mirabolfathi}, \citenamefont {Harris}, \citenamefont {Mahapatra},
  \citenamefont {Sundqvist}, \citenamefont {Jastram}, \citenamefont {Serfass},
  \citenamefont {Faiez},\ and\ \citenamefont {Sadoulet}}]{Contactfree}%
  \BibitemOpen
  \bibfield  {author} {\bibinfo {author} {\bibfnamefont {N.}~\bibnamefont
  {Mirabolfathi}}, \bibinfo {author} {\bibfnamefont {H.~R.}\ \bibnamefont
  {Harris}}, \bibinfo {author} {\bibfnamefont {R.}~\bibnamefont {Mahapatra}},
  \bibinfo {author} {\bibfnamefont {K.}~\bibnamefont {Sundqvist}}, \bibinfo
  {author} {\bibfnamefont {A.}~\bibnamefont {Jastram}}, \bibinfo {author}
  {\bibfnamefont {B.}~\bibnamefont {Serfass}}, \bibinfo {author} {\bibfnamefont
  {D.}~\bibnamefont {Faiez}}, \ and\ \bibinfo {author} {\bibfnamefont
  {B.}~\bibnamefont {Sadoulet}},\ }\href {\doibase 10.1016/j.nima.2017.02.032}
  {\bibfield  {journal} {\bibinfo  {journal} {Nucl. Instrum. Meth.}\ }\textbf
  {\bibinfo {volume} {A855}},\ \bibinfo {pages} {88} (\bibinfo {year}
  {2017})},\ \Eprint {http://arxiv.org/abs/1510.00999} {arXiv:1510.00999
  [physics.ins-det]} \BibitemShut {NoStop}%
\bibitem [{\citenamefont {Valdes}\ \emph {et~al.}(2003)\citenamefont {Valdes},
  \citenamefont {Parra}, \citenamefont {Diaz-Valdes}, \citenamefont {Denton},
  \citenamefont {Agurto}, \citenamefont {Ortega}, \citenamefont {Arista},\ and\
  \citenamefont {Vargas}}]{Val03}%
  \BibitemOpen
  \bibfield  {author} {\bibinfo {author} {\bibfnamefont {J.~E.}\ \bibnamefont
  {Valdes}}, \bibinfo {author} {\bibfnamefont {C.}~\bibnamefont {Parra}},
  \bibinfo {author} {\bibfnamefont {J.}~\bibnamefont {Diaz-Valdes}}, \bibinfo
  {author} {\bibfnamefont {C.~D.}\ \bibnamefont {Denton}}, \bibinfo {author}
  {\bibfnamefont {C.}~\bibnamefont {Agurto}}, \bibinfo {author} {\bibfnamefont
  {F.}~\bibnamefont {Ortega}}, \bibinfo {author} {\bibfnamefont {N.~R.}\
  \bibnamefont {Arista}}, \ and\ \bibinfo {author} {\bibfnamefont
  {P.}~\bibnamefont {Vargas}},\ }\href {\doibase 10.1103/PhysRevA.68.064901}
  {\bibfield  {journal} {\bibinfo  {journal} {Phys. Rev. A}\ }\textbf {\bibinfo
  {volume} {68}},\ \bibinfo {pages} {064901} (\bibinfo {year}
  {2003})}\BibitemShut {NoStop}%
\bibitem [{\citenamefont {Markin}\ \emph {et~al.}(2009)\citenamefont {Markin},
  \citenamefont {Primetzhofer},\ and\ \citenamefont {Bauer}}]{Mar09}%
  \BibitemOpen
  \bibfield  {author} {\bibinfo {author} {\bibfnamefont {S.~N.}\ \bibnamefont
  {Markin}}, \bibinfo {author} {\bibfnamefont {D.}~\bibnamefont
  {Primetzhofer}}, \ and\ \bibinfo {author} {\bibfnamefont {P.}~\bibnamefont
  {Bauer}},\ }\href@noop {} {\bibfield  {journal} {\bibinfo  {journal}
  {{Physical Review Letters}}\ }\textbf {\bibinfo {volume} {{103}}},\ \bibinfo
  {pages} {113201} (\bibinfo {year} {{2009}})}\BibitemShut {NoStop}%
\bibitem [{\citenamefont {Primetzhofer}(2012)}]{Pri12}%
  \BibitemOpen
  \bibfield  {author} {\bibinfo {author} {\bibfnamefont {D.}~\bibnamefont
  {Primetzhofer}},\ }\href@noop {} {\bibfield  {journal} {\bibinfo  {journal}
  {Phys. Rev. B}\ }\textbf {\bibinfo {volume} {86}},\ \bibinfo {pages} {094102}
  (\bibinfo {year} {2012})}\BibitemShut {NoStop}%
\bibitem [{\citenamefont {Lim}\ \emph {et~al.}(2016)\citenamefont {Lim},
  \citenamefont {Foulkes}, \citenamefont {Horsfield}, \citenamefont {Mason},
  \citenamefont {Schleife}, \citenamefont {Draeger},\ and\ \citenamefont
  {Correa}}]{Lim16}%
  \BibitemOpen
  \bibfield  {author} {\bibinfo {author} {\bibfnamefont {A.}~\bibnamefont
  {Lim}}, \bibinfo {author} {\bibfnamefont {W.~M.~C.}\ \bibnamefont {Foulkes}},
  \bibinfo {author} {\bibfnamefont {A.~P.}\ \bibnamefont {Horsfield}}, \bibinfo
  {author} {\bibfnamefont {D.~R.}\ \bibnamefont {Mason}}, \bibinfo {author}
  {\bibfnamefont {A.}~\bibnamefont {Schleife}}, \bibinfo {author}
  {\bibfnamefont {E.~W.}\ \bibnamefont {Draeger}}, \ and\ \bibinfo {author}
  {\bibfnamefont {A.~A.}\ \bibnamefont {Correa}},\ }\href@noop {} {\bibfield
  {journal} {\bibinfo  {journal} {Phys. Rev. Lett.}\ }\textbf {\bibinfo
  {volume} {116}},\ \bibinfo {pages} {043201} (\bibinfo {year}
  {2016})}\BibitemShut {NoStop}%
\bibitem [{\citenamefont {Horsfield}\ \emph {et~al.}(2016)\citenamefont
  {Horsfield}, \citenamefont {Lim}, \citenamefont {Foulkes},\ and\
  \citenamefont {Correa}}]{Hor16}%
  \BibitemOpen
  \bibfield  {author} {\bibinfo {author} {\bibfnamefont {A.~P.}\ \bibnamefont
  {Horsfield}}, \bibinfo {author} {\bibfnamefont {A.}~\bibnamefont {Lim}},
  \bibinfo {author} {\bibfnamefont {W.~M.~C.}\ \bibnamefont {Foulkes}}, \ and\
  \bibinfo {author} {\bibfnamefont {A.~A.}\ \bibnamefont {Correa}},\ }\href
  {\doibase 10.1103/PhysRevB.93.245106} {\bibfield  {journal} {\bibinfo
  {journal} {Phys. Rev. B}\ }\textbf {\bibinfo {volume} {93}},\ \bibinfo
  {pages} {245106} (\bibinfo {year} {2016})}\BibitemShut {NoStop}%
\bibitem [{\citenamefont {Holmstr{\"o}m}\ \emph {et~al.}(2008)\citenamefont
  {Holmstr{\"o}m}, \citenamefont {Kuronen},\ and\ \citenamefont
  {Nordlund}}]{Hol08a}%
  \BibitemOpen
  \bibfield  {author} {\bibinfo {author} {\bibfnamefont {E.}~\bibnamefont
  {Holmstr{\"o}m}}, \bibinfo {author} {\bibfnamefont {A.}~\bibnamefont
  {Kuronen}}, \ and\ \bibinfo {author} {\bibfnamefont {K.}~\bibnamefont
  {Nordlund}},\ }\href@noop {} {\bibfield  {journal} {\bibinfo  {journal}
  {Phys. Rev. B}\ }\textbf {\bibinfo {volume} {78}},\ \bibinfo {pages} {045202}
  (\bibinfo {year} {2008})}\BibitemShut {NoStop}%
\bibitem [{\citenamefont {Holmström}\ \emph {et~al.}(2010)\citenamefont
  {Holmström}, \citenamefont {Nordlund},\ and\ \citenamefont
  {Kuronen}}]{Hol10a}%
  \BibitemOpen
  \bibfield  {author} {\bibinfo {author} {\bibfnamefont {E.}~\bibnamefont
  {Holmström}}, \bibinfo {author} {\bibfnamefont {K.}~\bibnamefont
  {Nordlund}}, \ and\ \bibinfo {author} {\bibfnamefont {A.}~\bibnamefont
  {Kuronen}},\ }\href@noop {} {\bibfield  {journal} {\bibinfo  {journal}
  {Physica Scripta}\ }\textbf {\bibinfo {volume} {81}},\ \bibinfo {pages}
  {035601} (\bibinfo {year} {2010})}\BibitemShut {NoStop}%
\bibitem [{\citenamefont {Nordlund}\ \emph {et~al.}(2005)\citenamefont
  {Nordlund}, \citenamefont {Wallenius},\ and\ \citenamefont
  {Malerba}}]{Nor05c}%
  \BibitemOpen
  \bibfield  {author} {\bibinfo {author} {\bibfnamefont {K.}~\bibnamefont
  {Nordlund}}, \bibinfo {author} {\bibfnamefont {J.}~\bibnamefont {Wallenius}},
  \ and\ \bibinfo {author} {\bibfnamefont {L.}~\bibnamefont {Malerba}},\
  }\href@noop {} {\bibfield  {journal} {\bibinfo  {journal} {Nucl. Instr. Meth.
  Phys. Res. B}\ }\textbf {\bibinfo {volume} {246}},\ \bibinfo {pages} {322}
  (\bibinfo {year} {2005})}\BibitemShut {NoStop}%
\bibitem [{\citenamefont {Allen}\ and\ \citenamefont
  {Tildesley}(1989)}]{Allen-Tildesley}%
  \BibitemOpen
  \bibfield  {author} {\bibinfo {author} {\bibfnamefont {M.~P.}\ \bibnamefont
  {Allen}}\ and\ \bibinfo {author} {\bibfnamefont {D.~J.}\ \bibnamefont
  {Tildesley}},\ }\href@noop {} {\emph {\bibinfo {title} {{Computer Simulation
  of Liquids}}}}\ (\bibinfo  {publisher} {Oxford University Press},\ \bibinfo
  {address} {Oxford, England},\ \bibinfo {year} {1989})\BibitemShut {NoStop}%
\bibitem [{\citenamefont {Ding}\ and\ \citenamefont {Andersen}(1986)}]{Din86}%
  \BibitemOpen
  \bibfield  {author} {\bibinfo {author} {\bibfnamefont {K.}~\bibnamefont
  {Ding}}\ and\ \bibinfo {author} {\bibfnamefont {H.~C.}\ \bibnamefont
  {Andersen}},\ }\href@noop {} {\bibfield  {journal} {\bibinfo  {journal}
  {Phys. Rev. B}\ }\textbf {\bibinfo {volume} {34}},\ \bibinfo {pages} {6987}
  (\bibinfo {year} {1986})}\BibitemShut {NoStop}%
\bibitem [{\citenamefont {Nordlund}\ \emph {et~al.}(1998)\citenamefont
  {Nordlund}, \citenamefont {Ghaly}, \citenamefont {Averback}, \citenamefont
  {Caturla}, \citenamefont {{Diaz de la Rubia}},\ and\ \citenamefont
  {Tarus}}]{Nor97f}%
  \BibitemOpen
  \bibfield  {author} {\bibinfo {author} {\bibfnamefont {K.}~\bibnamefont
  {Nordlund}}, \bibinfo {author} {\bibfnamefont {M.}~\bibnamefont {Ghaly}},
  \bibinfo {author} {\bibfnamefont {R.~S.}\ \bibnamefont {Averback}}, \bibinfo
  {author} {\bibfnamefont {M.}~\bibnamefont {Caturla}}, \bibinfo {author}
  {\bibfnamefont {T.}~\bibnamefont {{Diaz de la Rubia}}}, \ and\ \bibinfo
  {author} {\bibfnamefont {J.}~\bibnamefont {Tarus}},\ }\href@noop {}
  {\bibfield  {journal} {\bibinfo  {journal} {Phys. Rev. B}\ }\textbf {\bibinfo
  {volume} {57}},\ \bibinfo {pages} {7556} (\bibinfo {year}
  {1998})}\BibitemShut {NoStop}%
\bibitem [{\citenamefont {Posselt}\ and\ \citenamefont
  {Gabriel}(2009)}]{Pos09}%
  \BibitemOpen
  \bibfield  {author} {\bibinfo {author} {\bibfnamefont {M.}~\bibnamefont
  {Posselt}}\ and\ \bibinfo {author} {\bibfnamefont {A.}~\bibnamefont
  {Gabriel}},\ }\href@noop {} {\bibfield  {journal} {\bibinfo  {journal} {Phys.
  Rev. B}\ }\textbf {\bibinfo {volume} {80}},\ \bibinfo {pages} {045202}
  (\bibinfo {year} {2009})}\BibitemShut {NoStop}%
\bibitem [{\citenamefont {Stillinger}\ and\ \citenamefont
  {Weber}(1985)}]{Sti85}%
  \BibitemOpen
  \bibfield  {author} {\bibinfo {author} {\bibfnamefont {F.~H.}\ \bibnamefont
  {Stillinger}}\ and\ \bibinfo {author} {\bibfnamefont {T.~A.}\ \bibnamefont
  {Weber}},\ }\href@noop {} {\bibfield  {journal} {\bibinfo  {journal} {Phys.
  Rev. B}\ }\textbf {\bibinfo {volume} {31}},\ \bibinfo {pages} {5262}
  (\bibinfo {year} {1985})}\BibitemShut {NoStop}%
\bibitem [{\citenamefont {O'Hare}\ \emph {et~al.}(2015)\citenamefont {O'Hare},
  \citenamefont {Green}, \citenamefont {Billard}, \citenamefont
  {Figueroa-Feliciano},\ and\ \citenamefont {Strigari}}]{rate}%
  \BibitemOpen
  \bibfield  {author} {\bibinfo {author} {\bibfnamefont {C.~A.~J.}\
  \bibnamefont {O'Hare}}, \bibinfo {author} {\bibfnamefont {A.~M.}\
  \bibnamefont {Green}}, \bibinfo {author} {\bibfnamefont {J.}~\bibnamefont
  {Billard}}, \bibinfo {author} {\bibfnamefont {E.}~\bibnamefont
  {Figueroa-Feliciano}}, \ and\ \bibinfo {author} {\bibfnamefont {L.~E.}\
  \bibnamefont {Strigari}},\ }\href {\doibase 10.1103/PhysRevD.92.063518}
  {\bibfield  {journal} {\bibinfo  {journal} {Phys. Rev.}\ }\textbf {\bibinfo
  {volume} {D92}},\ \bibinfo {pages} {063518} (\bibinfo {year} {2015})},\
  \Eprint {http://arxiv.org/abs/1505.08061} {arXiv:1505.08061 [astro-ph.CO]}
  \BibitemShut {NoStop}%
\bibitem [{\citenamefont {Duda}\ \emph {et~al.}(2007)\citenamefont {Duda},
  \citenamefont {Kemper},\ and\ \citenamefont {Gondolo}}]{fsq}%
  \BibitemOpen
  \bibfield  {author} {\bibinfo {author} {\bibfnamefont {G.}~\bibnamefont
  {Duda}}, \bibinfo {author} {\bibfnamefont {A.}~\bibnamefont {Kemper}}, \ and\
  \bibinfo {author} {\bibfnamefont {P.}~\bibnamefont {Gondolo}},\ }\href
  {\doibase 10.1088/1475-7516/2007/04/012} {\bibfield  {journal} {\bibinfo
  {journal} {JCAP}\ }\textbf {\bibinfo {volume} {0704}},\ \bibinfo {pages}
  {012} (\bibinfo {year} {2007})},\ \Eprint
  {http://arxiv.org/abs/hep-ph/0608035} {arXiv:hep-ph/0608035 [hep-ph]}
  \BibitemShut {NoStop}%
\bibitem [{\citenamefont {Bozorgnia}\ \emph {et~al.}(2011)\citenamefont
  {Bozorgnia}, \citenamefont {Gelmini},\ and\ \citenamefont {Gondolo}}]{vel}%
  \BibitemOpen
  \bibfield  {author} {\bibinfo {author} {\bibfnamefont {N.}~\bibnamefont
  {Bozorgnia}}, \bibinfo {author} {\bibfnamefont {G.~B.}\ \bibnamefont
  {Gelmini}}, \ and\ \bibinfo {author} {\bibfnamefont {P.}~\bibnamefont
  {Gondolo}},\ }\href {\doibase 10.1103/PhysRevD.84.023516} {\bibfield
  {journal} {\bibinfo  {journal} {Phys. Rev.}\ }\textbf {\bibinfo {volume}
  {D84}},\ \bibinfo {pages} {023516} (\bibinfo {year} {2011})},\ \Eprint
  {http://arxiv.org/abs/1101.2876} {arXiv:1101.2876 [astro-ph.CO]} \BibitemShut
  {NoStop}%
\bibitem [{Note1()}]{Note1}%
  \BibitemOpen
  \bibinfo {note} {Earth's revolution around the sun causes the annual
  modulation, whereas Earth's rotation causes the daily one.}\BibitemShut
  {Stop}%
\bibitem [{\citenamefont {Gorski}\ \emph {et~al.}(2005)\citenamefont {Gorski},
  \citenamefont {Hivon}, \citenamefont {Banday}, \citenamefont {Wandelt},
  \citenamefont {Hansen}, \citenamefont {Reinecke},\ and\ \citenamefont
  {Bartelman}}]{hp}%
  \BibitemOpen
  \bibfield  {author} {\bibinfo {author} {\bibfnamefont {K.~M.}\ \bibnamefont
  {Gorski}}, \bibinfo {author} {\bibfnamefont {E.}~\bibnamefont {Hivon}},
  \bibinfo {author} {\bibfnamefont {A.~J.}\ \bibnamefont {Banday}}, \bibinfo
  {author} {\bibfnamefont {B.~D.}\ \bibnamefont {Wandelt}}, \bibinfo {author}
  {\bibfnamefont {F.~K.}\ \bibnamefont {Hansen}}, \bibinfo {author}
  {\bibfnamefont {M.}~\bibnamefont {Reinecke}}, \ and\ \bibinfo {author}
  {\bibfnamefont {M.}~\bibnamefont {Bartelman}},\ }\href {\doibase
  10.1086/427976} {\bibfield  {journal} {\bibinfo  {journal} {Astrophys. J.}\
  }\textbf {\bibinfo {volume} {622}},\ \bibinfo {pages} {759} (\bibinfo {year}
  {2005})},\ \Eprint {http://arxiv.org/abs/astro-ph/0409513}
  {arXiv:astro-ph/0409513 [astro-ph]} \BibitemShut {NoStop}%
\bibitem [{\citenamefont {Gemmell}(1974)}]{Gemmell:1974ub}%
  \BibitemOpen
  \bibfield  {author} {\bibinfo {author} {\bibfnamefont {D.~S.}\ \bibnamefont
  {Gemmell}},\ }\href {\doibase 10.1103/RevModPhys.46.129} {\bibfield
  {journal} {\bibinfo  {journal} {Rev. Mod. Phys.}\ }\textbf {\bibinfo {volume}
  {46}},\ \bibinfo {pages} {129} (\bibinfo {year} {1974})}\BibitemShut
  {NoStop}%
\bibitem [{\citenamefont {Blackman}(1955)}]{Debyedisplacements}%
  \BibitemOpen
  \bibfield  {author} {\bibinfo {author} {\bibfnamefont {M.}~\bibnamefont
  {Blackman}},\ }\href@noop {} {\emph {\bibinfo {title} {{Handbuch der
  Physik}}}},\ Vol.\ \bibinfo {volume} {VII}\ (\bibinfo  {publisher}
  {Springer-Verlag},\ \bibinfo {address} {Berlin},\ \bibinfo {year}
  {1955})\BibitemShut {NoStop}%
\end{thebibliography}%

\end{document}